\documentstyle[preprint,tighten,aps,psfig]{revtex}

\begin{document}
%\twocolumn

\preprint{BNL-HET-98/12, TTP98-14, hep-ph/9804215}
\draft

\date{March 1998}
\title{Two-loop QCD corrections to semileptonic \boldmath $b$ decays at an
 intermediate recoil} 

\author{Andrzej Czarnecki}
\address{Physics Department, Brookhaven National Laboratory,\\
Upton, NY 11973}

\author{Kirill Melnikov}
\address{Institut f\"ur Theoretische Teilchenphysik, 
Universit\"at Karlsruhe,\\
D-76128 Karlsruhe, Germany}
\maketitle

\begin{abstract}
We present complete ${\cal O}(\alpha_s^2)$ corrections to the decay $b
\to c l \nu_l$ at the point where the invariant mass of the leptons
$\sqrt{q^2}$ equals the $c$ quark mass.  We use this result, together
with previously obtained corrections at the ends of the $q^2$
spectrum, to estimate the total width of the  semileptonic  $b\to c$
decay with ${\cal O}(\alpha_s^2)$ accuracy, essential for the
$|V_{cb}|$ determination.
\end{abstract}

\vspace*{3mm}

Precise determination of the Cabibbo-Kobayashi-Maskawa (CKM) matrix
parameters is one of the main goals of the present and upcoming high
energy experiments.  One of the three parameters which can be measured
with the highest accuracy is $|V_{cb}|$, for which there are two
complementary methods (for a recent review, see
e.g. \cite{Bigi97,Neubert97}).  The first one is based on studying the
exclusive decay $B\to D^\star + \mbox{leptons}$ near the zero recoil
end of the phase space.  The other, so-called inclusive method
examines the total semileptonic width of the $B$ meson, $\Gamma_{\rm
sl}(B \to X_c l \nu_l)$.  For both determinations the experimental
precision has reached a few per cent level and will be further
improved in the future $B$-factories.  To fully exploit those
measurements, the theoretical prediction for the $b$ quark
semileptonic width must be known with comparable precision.  This
requires a study of both short-distance (perturbative) and
long-distance (non-perturbative) QCD effects.  For the exclusive
method the ${\cal O}(\alpha_s^2)$ corrections were calculated in
\cite{CzMeUr}. Technically, the most challenging part of that
calculation was the QCD correction to the axial $b \to c$ current
normalization at zero recoil, obtained in
\cite{zerorecoil,zerorecoilA} and confirmed in a recent study
\cite{Franzkowski:1997vg}.  Theoretical accuracy of the exclusive
method is limited by the errors in the non--perturbative matrix
elements which are enhanced due to the not so large mass of the $c$
quark.

In the inclusive method the non--perturbative corrections are somewhat
smaller, with suppression by $1/m_b^2$. They are estimated
\cite{Bigi:1992su} to decrease the meson semileptonic width by
approximately 5\% compared to the free quark decay rate.  With the
non--perturbative corrections under control, the perturbative
corrections should be carefully analyzed.  Ideally one would like to
know the two loop QCD correction for an arbitrary invariant mass
squared of the leptons ($q^2$) in the decay $b\to cl\nu_l$.  This has
been achieved for only a subset of corrections, enhanced by a large
factor describing the running of the strong coupling constant
\cite{Luke:1995}; these so-called Brodsky-Lepage-Mackenzie (BLM)
corrections \cite{BLM} have been even resummed to all orders
\cite{Ball:1995}.  However, in absence of a complete calculation of
the two-loop QCD effects, the related uncertainty in $|V_{cb}|$
determined from the inclusive method had to be guessed and various
estimates were given in the literature. It is clear that the problem
can be only solved by an explicit calculation of the complete ${\cal
O}(\alpha_s^2)$ effects and therefore the need for such a calculation
was often emphasized.  In the present paper we demonstrate that, with
the new result presented here, perturbative effects can be estimated
with sufficient accuracy.

For an arbitrary value of $q^2$ (and, by the same token, for the total
width), the calculation of the second order corrections is a very
difficult technical task.  In earlier works we presented such results
for the two extreme points, $q^2=0$ \cite{Czarnecki:1997hc} and
$q^2=(m_b-m_c)^2$ \cite{zerorecoil,zerorecoilA}.  While at both points
the non-BLM corrections were below 1\% of the total semileptonic
width, it was not clear what happens at intermediate values of $q^2$,
where a complete calculation appeared infeasible.  Nevertheless, from
the results in zero and maximal recoil points it was conjectured in
\cite{Czarnecki:1997hc}, that the second order non--BLM corrections to
the semileptonic decay width $b\to cl\nu_l$ do not exceed $1\%$.

\begin{figure} 
\hspace*{-36mm}
\begin{minipage}{16.cm}
\vspace*{3mm}
\[
\hspace*{70mm}
\mbox{
\psfig{figure=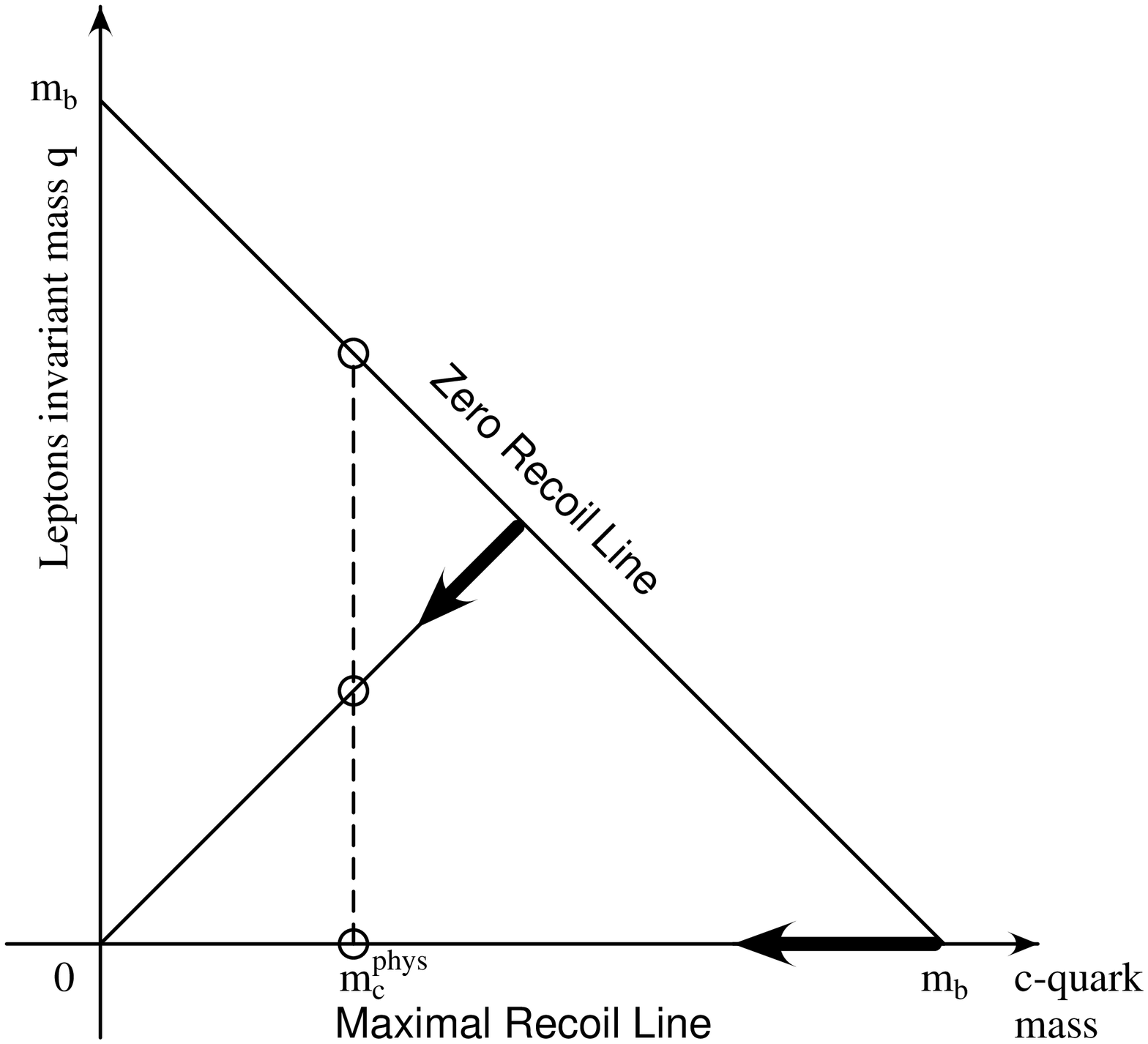,width=80mm,bbllx=25pt,bblly=104pt,%
bburx=582pt,bbury=630pt} 
}
\]
\end{minipage}
\caption{Status of the two-loop QCD corrections to the decay $b\to
c+\mbox{leptons}$.  The dashed line denotes the physical region for
the actual $c$ quark.  Points where the full corrections are known are
circled.  An analytical formula is known along the whole zero recoil
line \protect\cite{zerorecoilA,Franzkowski:1997vg}.  The other two
points are found from expansions from the base points of the two
arrows: at maximal recoil \protect\cite{Czarnecki:1997hc} and at the
intersection with the diagonal (present work).}
\label{fig:triangle}
\end{figure}

Since then, we have found another kinematical point, $q^2 = m_c^2$,
where the complete calculation of the second order QCD corrections is
possible. It is fortunate that for the physical value of $m_c$ this
roughly corresponds to the middle of the $q$ range, so that together
with the endpoints we have nicely distributed information (see
fig.~\ref{fig:triangle}).

Thus, the purpose of this paper is twofold. We first describe a
calculation of the ${\cal O}(\alpha_s^2)$ correction to the
differential semileptonic width of the $b$ quark for $q^2 = m_c^2$.
We then use the results obtained in this paper, as well as in
\cite{zerorecoil,zerorecoilA,Czarnecki:1997hc}, to estimate the ${\cal
O}(\alpha_s^2)$ correction  to the semileptonic decay width of the
$b$--quark.
 
We start with a short description of the new calculation of the ${\cal
O}(\alpha_s^2)$ correction for $q^2 = m_c^2$. A complete description
of the technical details of the present calculation will be presented
elsewhere. The basic idea is the expansion of the decay amplitude in
the velocity of the massive quark in the final state.

The Feynman diagrams which describe this process can be divided up
into three classes according to the number of real gluons emitted.  In
the first group there are the purely virtual two-loop corrections.
Their value is known at the point $m_c=q=m_b/2$ from our previous
study on the zero recoil line \cite{zerorecoilA}.  Since the actual
value of $m_c$ is not equal $m_b/2$, we expand the diagrams in the
parameter $\beta$, related to the difference of these two quantities
(see below).  The coefficients of this expansion are given by diagrams
with higher powers of propagators.  By solving a system of recurrence
relations, obtained by integration by parts, we found an algorithm
with which all these Feynman integrals can be expressed in terms of
four non-trivial functions, given below in eq.~(\ref{eq:fs}), and
some simpler (single--scale) Feynman integrals.  This is the most
challenging part of this calculation and it limits the number of terms
we have been able to obtain using the present computing facilities.

In the other two groups there are diagrams with two real gluons or one
real gluon and a virtual loop. It is relatively easy to perform
calculations in these cases using (slightly modified)
computational techniques described in detail in \cite{maxtech}.  This
possibility is related to the fact that for $q^2 = m_c^2$ the three--
and four--particle phase space integrals can be expressed
in terms of the Euler $\Gamma$ function.

We write the differential semileptonic decay width of the decay $b\to
c l \nu$ at $q^2=m_c^2$ as
\begin{eqnarray}
\left[ \frac {{\rm d}\Gamma_{\rm sl}}{{\rm d} q^2} \right]_{q^2=m_c^2}
= {G_F^2 m_b^3\over 96\pi^3}|V_{cb}|^2
\left[
\Delta_{\rm Born}
+ {\alpha_s\over\pi} C_F \Delta_1
+ \left({\alpha_s\over\pi}\right)^2 C_F \Delta_2
\right],
\end{eqnarray}
where $\Delta_{{\rm Born},1,2}$ describe the $m_c/m_b$ dependence in
various orders in the strong coupling constant.  Both $\Delta_{{\rm
Born}} = \left(1-m_c^2/m_b^2\right) \sqrt{1-4m_c^2/m_b^2}$ and
$\Delta_1$ are known in a closed analytical form \cite{jk2,cza90}.
$\Delta_2$ is calculated in the present paper. For the purpose of
presentation we divide it up into four contributions according to the
color factors:
\begin{eqnarray}
\Delta_2 &=& \sqrt{1-4m_c^2/m_b^2}~~\left[ 
(C_F-C_A/2)\Delta_{F}
+C_A\Delta_{A}
+T_RN_L \Delta_{L}
+T_R \Delta_{H}
\right].
\label{eq:delta2}
\end{eqnarray}
The last term, $\Delta_{H}$, describes the contributions of the
massive $b$ and $c$ quark loops.  Top quark contribution is suppressed
by a factor $\sim m_b^2/m_t^2$ and has been neglected.  For the SU(3)
group the color factors are $C_A=3$, $C_F=4/3$, $T_R=1/2$.  $N_L=3$ is
the number of the quark flavors whose masses have been neglected ($u$,
$d$, and $s$).

We have computed 5 terms of the expansion in $\beta \equiv
(1-4m_c^2/m_b^2)$ of $\Delta_{F,A,L,H}$.  In order to save space, in
the formulas presented here we give only numerical values of the
coefficients of the second and higher powers of $\beta$ (these
coefficients contain $\ln(\beta)$; we use $\beta=0.64$ to evaluate
them).  We use the $\overline{MS}$ scheme for the strong coupling
constant and normalize it at the scale $\sqrt{m_b m_c}$, which has
been  proposed as optimal for
the semileptonic $b \to c$ transitions, in view of the proximity of
the small velocity limit \cite{Bigi:1997si}.  Using the pole mass of the
$b$ and $c$ quarks and expressing the one-loop corrections in terms of
$\alpha_s(\sqrt{m_b m_c})$ we find
\begin{eqnarray}
\Delta_{A}  &=&
     {49 \over 96} - {19 \over 96} \pi^2 \ln(2) + {121 \over 384}
       \pi^2 - {591 \over 128} \ln(2) + {179 \over 64} \ln^2(2)
 - {51\over 64} \zeta_3 
\nonumber \\ &&
       + \beta \left[  - {89669 \over 6912} - {17 \over 288} \pi^2 \ln(2)
       - {1 \over 12} \pi^2 \ln(\beta) + {43 \over 96} \pi^2 
          - {7 \over 6} \ln(2) \ln(\beta) + {10225 \over 1152} \ln(2)
\right.  \nonumber \\ && \left. \qquad
 +       {523 \over 192} \ln^2(2)
 + {89 \over 64} \zeta_3 + {253\over  
         72} \ln(\beta) - {11 \over 24} \ln^2(\beta) \right]
\nonumber \\ &&
       - 2.2631\beta^2 -    1.1747 \beta^3 - 0.3171 \beta^4,
  \nonumber \\ 
%%%%%%%%%%%%%%%%%%%%%%%%%%%%%%%%%%%%%%%%%%%%%%%%%%
\Delta_{F}  &=& 
       {117 \over 16} - {19 \over 128} f_1 - {3 \over 8} f_3 - {5
     \over 64} f_4 + {13 \over 12} \pi^2 \ln(2) - {173 \over 384}
     \pi^2 - {957 \over 64} \ln(2) + {409 \over 64} \ln^2(2)
  \nonumber \\ && \qquad
 - {57     \over 32} \zeta_3 - {291 \over 32} R_2  
\nonumber \\ &&
       + \beta   \left[  {313 \over 128} - {37 \over 192} f_1 - {1 \over 16}
     f_3 + {7 \over 96} f_4 + {55 \over 144} \pi^2 \ln(2) -  
         {6071 \over 27648} \pi^2 + {9 \over 4} \ln(2) \ln(\beta)
\right.  \nonumber \\ && \left. \qquad
 - {1825 \over 144} \ln(2) + {179813 \over 13824} \ln^2(2) 
 + {25 \over 32} \zeta_3 + {465 \over 256} R_2 - 2 \ln(\beta)  \right]
  \nonumber \\ && +  0.11065 \beta^2 + 
   0.54858 \beta^3 + 0.78239 \beta^4,
\nonumber \\ 
%%%%%%%%%%%%%%%%%%%%%%%%%%%%%%%%%%%%%%%%%%%%%%%%%%%%%%%%%%%%%%%%%
\Delta_{L}  &=& 
     {11 \over 12} - {15 \over 16} \ln(2) 
      \nonumber \\ && + \beta     \left[{2699 \over 864} - {1 \over 9}
     \pi^2 
     + {5 \over 6} 
\ln(2) \ln(\beta) - {517 \over 144} \ln(2) + \ln^2(2) -  
         {13 \over 9} \ln(\beta) + {1 \over 6} \ln^2(\beta)  \right]
       \nonumber \\ && +  0.72295 \beta^2 + 
   0.62301 \beta^3 + 0.26442 \beta^4,
\nonumber \\ 
%%%%%%%%%%%%%%%%%%%%%%%%%%%%%%%%%%%%%%%%%%%%%%%%%%%%%%%%%%%%%%%%%
\Delta_{H}  &=& 
 {1379 \over 96} + {355 \over 3072} \pi^2 + {195 \over 64}
     \ln(2) + {3317 \over 512} \ln^2(2) + {7371 \over 256} R_2  
  \nonumber \\ && + \beta  \left[  {1579 \over 216} + {1891 \over 1024}
     \pi^2 + {299 \over 192} \ln(2) + {6343 \over 512} \ln^2(2) +  
         {11385 \over 256} R_2   \right]
  \nonumber \\ &&  - 0.14557 \beta^2 - 0.075047 \beta^3 - 0.061437
 \beta^4. 
\label{eq:main}
\end{eqnarray}
In the above formulas we use the notations $f_{1,3,4}$ and $R_2$ for
the values at $\omega=1/2$ of so denoted functions, for which
analytical formulas are given in \cite{zerorecoilA}.  Numerically they
give
\begin{eqnarray}
f_1 &\approx &  3.24460, \qquad f_3 \approx 12.3201,
\nonumber \\
f_4 &\approx &  7.83195, \qquad R_2 \approx -0.72946.
\label{eq:fs}
\end{eqnarray}

We now would like to estimate the uncalculated remainder of the series
given in (\ref{eq:main}) and the error in the final result.  In a
series $\sum a_n \beta^n$ the remainder is less than the last known
term multiplied by $\beta/(1-\beta)$, provided that $a_n$ is a
decreasing sequence, which we assume here.  Therefore we estimate the
final result by adding to the known terms half of $\beta/(1-\beta)$
times the last term.  This additional term also gives a conservative
estimate of the error.  This procedure does not apply directly to
$\Delta_F$, where the coefficients appear to grow.  From previous
experience with similar calculations we think that this is because the
series describing $\Delta_F$ has not yet achieved its asymptotic
behavior $\sim 1/n$.  For example, in the maximal recoil case
\cite{maxtech} the first five terms of the expansion for $\Delta_F$
behave rather wildly; nevertheless, they approximate the final result
with an accuracy of about 25\% (fortunately, the final result for
$\Delta_2$ is rather insensitive to this error).  We assign a similar
error bar to $\Delta_F$.  For $\beta=0.64$ we find:
\begin{eqnarray}
\Delta_{A}  &=&-2.37(5),\qquad \Delta_{F}  = 1.78(50),
\nonumber \\
\Delta_{L}  &=& 1.09(4), \qquad \Delta_{H}  = -0.29(1).
\end{eqnarray}
Finally, we get for the correction defined in
eq.~(\ref{eq:delta2}) (we add the errors in quadrature)
\begin{equation}
\Delta_2=  -4.72(14).
\label{eq:del2}
\end{equation}
We now summarize the information about perturbative corrections to $b
\to c$ transitions and estimate the ${\cal O}(\alpha_s^2)$ correction
to the total semileptonic decay width of the $b$--quark.  We use the
results presented in this paper, as well as in our previous papers 
\cite{zerorecoil,zerorecoilA,Czarnecki:1997hc}.  For arbitrary $q^2$,
we define:
\begin{eqnarray}
 \frac {{\rm d}\Gamma_{\rm sl}}{{\rm d} q^2} 
= {G_F^2 m_b^3\over 96\pi^3}|V_{cb}|^2
\left[
A_{\rm Born}
+ {\alpha_s (\sqrt{m_b m_c})\over\pi} C_F A_1
+ \left({\alpha_s\over\pi}\right)^2 C_F A_2
\right]
\end{eqnarray}
where 
$m_b$ refers to the pole mass of the $b$ quark.

The BLM part of $A_2$ was obtained in \cite{Luke:1995}.
The difference between
the complete $A_2$ and $A_2^{\rm BLM}$ is  called
a non--BLM correction.  We introduce a quantity $\xi$,
\begin{equation}
\xi (q^2) = \frac {A_2 (q^2) - A_2^{\rm BLM}(q^2)}
{A_{\rm Born}(q^2)},
\label{10}
\end{equation}
which describes the size of the non--BLM correction relative to the
Born term as a function of $q^2$.  The values of $\xi (q^2)$, for
$q^2=0$ \cite{Czarnecki:1997hc}, $q^2 = m_c^2$ (this paper) and $q^2
=q_{\rm max}^2 \equiv (m_b-m_c)^2$ \cite{zerorecoil,zerorecoilA} are,
respectively, 0.65, 1.0, 0.06 (for $m_c/m_b=0.3$).

These three numbers permit us to estimate the ${\cal O}(\alpha_s^2)$
correction to the semileptonic decay width.  For this purpose, we
interpolate the non-BLM corrections with the function $\xi(q^2) = a_2
q^4 + a_1 q^2 + a_0$, where the coefficients $a_i$ are determined from
the above values of $\xi(0)$, $\xi(m_c^2)$, $\xi(q_{\rm max}^2)$.  The
function $\xi(q^2)$ is then integrated over $q^2$, using the Born
differential rate as the weight. As the result one gets an estimate of
the non--BLM correction for the total semileptonic decay rate:
\begin{equation}
\frac {\int \limits_{0}^{(m_b-m_c)^2} 
{\rm d} q^2 A_{\rm Born}(q^2) \xi(q^2)}
{ \int \limits_{0}^{(m_b-m_c)^2} 
{\rm d} q^2 A_{\rm Born}(q^2)}
\simeq 1.1.
\end{equation}

The validity of this estimate can be checked by using the same
procedure to obtain an estimate of the BLM correction for the total
decay rate using the known results at $q^2=0, m_c^2, (m_b-m_c)^2$ as
an input.

Such a fit results in the value of the BLM correction
$-8.6(\alpha_s/\pi)^2$, to be compared with the exact result
\cite{Luke:1995} $-9.8(\alpha_s/\pi)^2$.\footnote{Actually,
ref.~\protect\cite{Luke:1995} gives $-15.1(\alpha_s/\pi)^2$.  We
modify that value by using 4, rather than 3, massless flavors for the
$\alpha_s$ evolution, and by using $\sqrt{m_bm_c}$, rather than $m_b$,
as its normalization point.}  We conclude that the accuracy of our
simple fit is not worse than 30\%, which also includes the errors in
the three input data points.  We therefore obtain the following
formula for the total semileptonic decay rate $\Gamma_{\rm sl}(b \to
cl\nu_l)$:
\begin{equation}
\Gamma_{\rm sl} = {G_F^2 m_b^5\over 192\pi^3}|V_{cb}|^2  
F \left ( \frac {m_c^2}{m_b^2} \right )
\left[ 1 -1.67  \frac {\alpha_s (\sqrt{m_bm_c})}{\pi} 
+\left (-9.8 + 1.4\pm 0.4\right )  
\left ( \frac {\alpha_s}{\pi} \right )^2 \right],
\label{widthpole}
\end{equation}
where $F(x) = 1-8x-12x^2\ln(x) + 8x^3 - x^4$ and we used $m_c/m_b =
0.3$.  For the sake of clarity we separated the BLM and the non--BLM
parts of the second order corrections. We also explicitly indicated
the uncertainty in our estimate of the second order non--BLM
correction.

In principle, Eq.(\ref{widthpole}) provides the result for the
semileptonic decay width $b \to cl\nu_l$, when expressed through the
pole $b$ and $c$ quark masses, valid to ${\cal O}(\alpha_s ^2)$. One
notices that Eq.~(\ref{widthpole}) contains a large second order
correction due to the BLM effects. For a long time this observation
seemed to seriously limit the precision in $|V_{cb}|$, as obtained
from the inclusive method.  It was, however, noticed
\cite{Uraltsev95upset} that the large value of the second order BLM
corrections is correlated with the fact that the pole quark masses
were used in Eq.~(\ref{widthpole}).  It is well established, that the
pole quark masses can not be defined when non--perturbative
corrections are addressed. A hint that this is really the case is
given by the bad behavior of the perturbation series itself.  In the
case of the semileptonic $b$--decays the problem is enhanced, since
the decay width is proportional to the fifth power of the $b$--quark
mass.

It was suggested in \cite{Uraltsev95upset} and then further elaborated in
\cite{Bigi:1997si} that the most appropriate masses, to be used in the
expression for the decay width, are the so--called low--scale 
running quark
masses normalized at a scale $\mu \sim 1-2$ GeV.  On one hand, such
masses can be defined on the non--perturbative level, on the other
(and this is related to the first point) their use is expected to
minimize the perturbative corrections to the semileptonic decay width.
To ${\cal O}(\alpha_s^2)$ a (perturbative) relation between the pole
and the low--scale quark masses was obtained in \cite{CzMeUr1} and
reads:
\begin{equation}
m_{\rm pole} = m(\mu) + \left [\Lambda (\mu) \right ]_{\rm pert} +
\frac {1}{2 m(\mu)} \left [ \mu^2_\pi(\mu) \right ]_{\rm pert},
\label {polemass}
\end{equation}
where 
\begin{eqnarray}
\left [\Lambda (\mu) \right ]_{\rm pert} &=& 
\frac {4}{3}C_F\mu \frac {\alpha_s(\mu_0)}{\pi} \left \{
1 + \frac {\alpha_s}{\pi} \left [ 
\left (\frac {4}{3} - \frac {1}{2}\ln \frac {2\mu}{\mu _0} \right
)\beta_0 - C_A \left (\frac {\pi^2}{6} - \frac {13}{12} \right )
\right ] \right \},
\nonumber \\
\left [\mu_{\pi}^2 (\mu) \right ]_{\rm pert} &=& 
C_F\mu^2 \frac {\alpha_s(\mu _0)}{\pi} \left \{
1 + \frac {\alpha_s}{\pi} \left [ 
\left (\frac {13}{12} - \frac {1}{2} \ln \frac {2\mu}{\mu _0} \right
)\beta_0 - C_A \left (\frac {\pi^2}{6} - \frac {13}{12} \right )
\right ] \right \}.
\nonumber
\end{eqnarray}

It is important that, in contrast to the pole masses, the low--scale
masses do not have any significant numerical ambiguity.  We use
Eq.~(\ref{polemass}) to rewrite the expression for the semileptonic
decay width (\ref{widthpole}) through the low-scale masses normalized
at $\mu=1$ GeV. In the BLM approximation such calculation was
performed in \cite{Uraltsev95upset}.  As a result, we find that the
perturbative coefficients decrease noticeably:
\begin{eqnarray}
\Gamma_{\rm sl}(b \to c l \nu_l)  
&=& \frac {G_F^2 \widetilde{m}_b^5 |V_{cb}|^2}{192\pi^3} 
F\left ( \frac {\widetilde{m}_c^2}{\widetilde{m}_b^2} \right )
\left[ 1  
- 1.14  \frac {\alpha_s (\sqrt{\widetilde{m}_b \widetilde{m}_c})}{\pi} 
 -(2.65\pm 0.40)  \left ( \frac {\alpha_s}{\pi} \right )^2 
\right],
\label {Bigi:1992suole}
\end{eqnarray}
where we have used the values of the low scale running quark masses
$\widetilde{m}_b = 4.64(5)$ GeV and $\widetilde{m}_c = 1.25(10)$ GeV,
as suggested in \cite{Bigi97}. Though in \cite{Bigi97} the errors
assigned to the $b$ and $c$ quark  low scale running mass were
considered conservative, in our opinion this issue is not completely
clear and a dedicated analysis is required.  It is, however, rather
certain that in contrast to the pole mass, the accuracy of
$\widetilde{m}_Q$ can {\em in principle} be reliably estimated.

We see that the perturbative series for the inclusive width appears to
behave better when the low scale masses are used, in accordance with
the theoretical arguments \cite{Uraltsev95upset,Bigi:1997si,Bigi97}.

To sum, we have estimated  the  second order QCD corrections to the
width of the semileptonic $b\to c$ decay.  The small value of these
corrections shows that the perturbative series is not likely to cause
any significant uncertainty in the $|V_{cb}|$ extracted using the
inclusive method, provided that the decay width is calculated using
the low scale mass definition.  Further improvement in the theoretical
predictions for $\Gamma_{\rm sl}(B \to X_c l \nu_l)$ will be possible
when more precise quark mass values and the non--perturbative matrix
elements have been determined.

\vspace*{.3cm} We are indebted to N.G. Uraltsev for many helpful
comments.  We thank P.~Baikov for valuable discussions concerning the
solution of the recurrence relations for this problem, and J.B.~Tausk
for helping us check some Feynman integrals.  This work was supported
in part by the DOE contract DE-AC02-98CH10886 and by the grant
BMBF-057KA92P.

%\bibliographystyle{../tex/prsty}
%\bibliography{../tex/phd}

\end{document}